\documentclass{ws-procs9x6}

\usepackage{graphics}
\usepackage{psfig}
\usepackage{graphicx}
\usepackage{epsfig} 
\usepackage[usenames]{color}
\usepackage{dcolumn}

\newcommand{\ba}{\begin{eqnarray}}
\newcommand{\ea}{\end{eqnarray}}

\newcommand{\ov}{\overline}

\newcommand{\be}{\begin{equation}}
\newcommand{\ee}{\end{equation}}
\newcommand{\bea}{\begin{eqnarray}}
\newcommand{\eea}{\end{eqnarray}}
\newcommand{\pa}{\partial}

\newcommand{\bw}{\begin{widetext}}
\newcommand{\ew}{\end{widetext}}

\begin{document}

\title{Viscosity near phase transitions
\footnote{\uppercase{T}his work has been partially supported by the \uppercase{DGICYT} (\uppercase{S}pain) under grants \uppercase{FPA} 2008-00592, \uppercase{FIS}2008-01323 plus 227431, \uppercase{H}adron\uppercase{P}hysics2 (\uppercase{EU}) and
 \uppercase{PR}34-1856-\uppercase{BSCH}, \uppercase{UCM}-\uppercase{BSCH} \uppercase{GR}58/08, 910309, \uppercase{PR}34/07-15875. \uppercase{J. M. T.} is an \uppercase{FPU} scholarship recipient. \uppercase{A. D.} thanks \uppercase{J}ulia \uppercase{N}yiri for her kind invitation to
 participate in this celebration of \uppercase{G}ribov's work. 
}}
\author{Antonio Dobado, Felipe J. Llanes-Estrada and \\ Juan M. Torres-Rincon}
\address{Departamento de F\'isica Te\'orica I, Universidad Complutense de Madrid \\ 28040 Madrid, Spain 
}

\maketitle

\abstracts{
Probably the most enticing observation in theoretical physics during the last decade was the discovery of the great amount of consequences obtained from the AdS/CFT conjecture put forward by Maldacena. 
In this work we review how this correspondence can be used to address hydrodynamic properties such as the viscosity of
some strongly interacting systems. We also employ the Boltzmann equation for those systems closer to low-energy QCD, and argue that this kind of transport coefficients can be related to phase transitions, in particular the QGP/hadronic phase transition studied in heavy ion collisions.}

\section{Holography and the Maldacena conjecture}
One of the most successful tools developed in the last years in order to explore the strongly interacting regime of gauge theories is the idea of holography, or more properly the AdS/CFT
 (Anti de Sitter/Conformal Field Theory) correspondence introduced by Maldacena and refined by Witten more than one decade ago\cite{AdS}.  The correspondence states the equivalence 
of two theories defined in spaces of different dimension. In its original version the two theories were Type-IIB string theory  defined on $AdS_5\times S_5$ and $N=4$ supersymmetric $SU(N)$
 Yang-Mills (SYM)  theory defined in the four dimensional boundary of this space. Although this correspondence has not been proven so far, there is a vast body of evidence supporting it and thus
 it stands as a very strong conjecture that has also been generalized to many other cases. 

Let us try to understand the rationale of the conjecture in the following. The low energy excitations of a Type IIB string theory in the presence of a stack of $N$ 3D-branes can be seen in two very different ways. 
\begin{itemize}
\item The first description of the system reduces to a $N=4$ SYM theory in four dimensions (which is a conformal quantum field theory) plus free gravity in ten dimensions. 
\item The second way to look at the low energy excitations of the system is by thinking of an observer in the asymptotic, flat Minkowski region.  Then the only relevant modes are those belonging to the ten
 dimensional graviton multiplet. Near the branes the geometry becomes effectively $AdS_5 \times S_5$. Deep in this ``throat'' region an arbitrarily high-energy closed string (graviton) may appear to have an arbitrarily low
 energy as seen from an observer at the Minkowski infinity. Therefore we can conclude that the second description of the system is given by interacting closed strings in the mentioned $AdS_5 \times S_5$ background. 
\end{itemize}
This observation supports the idea that it is reasonable to conjecture that four dimensional  $N=4$ $SU(N)$ SYM theory and Type IIB string theory on $AdS_5 \times S_5$ are two different (dual) descriptions of
 the same physical system. It can also be checked that the symmetries of both theories are exactly the same which gives additional support to the conjecture.

In principle, the parameters needed to define the $N=4$ $SU(N)$ SYM theory (CFT in the following) are the gauge coupling $g$ and $N$. It is useful to define also the so called 't Hooft parameter $\lambda=g^2N$
 specially when dealing with the large $N$ limit. On the other hand, to define the string theory one needs to specify the string length $l_s$, the string coupling $g_{st}$ and the common size of the $AdS_5$
 and the $S_5$ spaces $R$. Then it is possible to relate the two sets of parameters by using the renowned ``Maldacena dictionary''
\begin{equation}
 g^2 = 4\pi g_{st}
\end{equation}
and
\begin{equation}
\lambda= g^2 N = \frac {R^4}{l_s^4}.                  
\end{equation}
This last equation is quite interesting. If one considers the strong coupling limit of the gauge theory (large t'Hooft coupling) we have also $l_s \ll R$, or in other words, we are in the region where
 string theory can be approximated by Supergravity (SUGRA) or even plain Einstein theory. Therefore it is possible to make reliable computations  in a strongly coupled theory by using its weakly coupled
 gravity dual. This highly non trivial result is one of the reasons for the enormous attention paid to the Maldacena conjecture. 

The holographic correspondence can also be extended to finite temperature $T > 0$. In order to do that one needs to introduce a black hole in the
AdS space. Then it is possible to find the well known Bekenstein-Hawking formulae relating
\begin{itemize}
  \item First the horizon area to the entropy $S$ and the horizon area $A$ viz. $S=A/4G$, $G$ being the Newton-Cavendish constant,

  \item and second, the temperature with the horizon radius $r_0$ and the whole space size,
    \begin{equation} \label{temperature} T=r_0 /\pi R^2. \end{equation}
\end{itemize}

In spite of the big theoretical appeal of the AdS/CFT ideas, it is fair to ask what is their real relevance for well established quantum field theories such as QCD. So far, no gravity dual has been found
 for this theory. However it is possible that the $N=4$ SYM theory could still bring some light about the QCD plasma studied at heavy ion colliders like RHIC. As we will see in the next sections, this
 is in particular the case of some transport coefficients like the shear viscosity.

\section{$\eta /s $ from the AdS/CFT correspondence}

One interesting (and little stressed) feature of hydrodynamics beyond its classical applications is that it can be used to describe the low energy (long wave) behavior of certain quantum field theories.
 The standard way to proceed is by expanding the energy-momentum tensor in terms of the number of space-time derivatives. At lowest order one gets the well known ideal fluid equations\cite{landau}.
The introduction of dissipative processes requires going to the next order where transport or diffusion coefficients such as the  shear viscosity $\eta$, the bulk viscosity $\zeta$ and the heat conductivity $\kappa$ appear,
together with other possible diffusion coefficients $D_i$ related with different conserved quantities such as flavors. 
The linearized equations of motion yields the dispersion relations corresponding for example to the two transverse modes 
\begin{equation} \label{transv}
\omega(k)=-i \ \frac{\eta}{\epsilon+P} \ k^2,
\end{equation}
(where $\epsilon$ is the energy density and $P$ the pressure) or  to the longitudinal sound mode:
\begin{equation}
\omega(k)=c_s k-\frac{i}{2}\left(\frac{4}{3}\eta+\zeta\right)\frac{k^2}{\epsilon+P},
\end{equation}
where $c_s=\sqrt{dP/d\epsilon}$ is the speed of sound. By using the thermodynamic relation $\epsilon+P=Ts$ valid for vanishing chemical potential, with $s$ being the entropy density, Eq.~(\ref{transv}) shows that $\eta/s$ characterizes the intrinsic ability of the system to relax towards equilibrium. Let us recall also that  $\zeta=0$ for scale-invariant theories.

The traditional way to compute transport coefficients is kinetic theory. The
Boltzmann equation or its quantum version, the
Uehling-Uhlenbenck equation, written in terms of the elastic
cross sections can be solved by the
Chapman-Enskog method to first order in the perturbation out of
equilibrium\cite{llanesdobado2003} and from the different perturbations one can find the different transport coefficients. However this method applies typically only for weakly-interacting, dilute enough systems.

A more modern way for computing transport coefficients is by using the Kubo formulae. By analyzing linear
response theory and coupling the system to gravity by using an
appropriate lightly curved space-time background it is possible to 
find:
\begin{equation}
\eta= \lim_{\omega\rightarrow 0}\left(\frac{1}
 {2\omega}\int dt d \bar x \
e^{i\omega t}\langle[T_{xy}(t,\bar x),T_{xy}(0,\bar 0) ]\rangle
\right) \ .
\end{equation}
One of the nice things about the Maldacena conjecture is that it is supposed to provide a complete description of the CFT even in the strong interaction regime. This makes possible to study the theory, 
not only at finite temperature, but also in the hydrodynamic regime. In particular the Kubo equation can be used to compute $\eta/s$ in the context of
the AdS/CFT correspondence. To do that one starts from a
CFT with gravity dual. For example for $N=4, SU(N)$ SYM one can consider the metric:
\begin{equation}
ds^2=\frac{r^2}{R^2}\left[-\left(1-\frac{r_0^4}{r^4}\right)dt^2+dx^2+dy^2+dz^2 \right]
+\frac{R^2}{r^2(1-r_0^4/r^4)}dr^2 \ .
\end{equation}
The dual theory is a CFT at temperature  $T$ given by Eq. (\ref{temperature}). Then it is possible to consider\cite{Klebanov:1997kc} a graviton polarized in the x-y direction propagating
perpendiculary to the brane. The absorption
cross section of the graviton by the brane measures, in the dual CFT,
the imaginary part of the retarded Green's function of the operator
coupled to the
metric i.e. the energy-momentum tensor. Then it is possible to
find:
\begin{equation}
\eta=\frac{\sigma(0)}{16\pi G},
\end{equation}
where $\sigma(0)$ is the graviton absorption cross section at zero
energy. This cross section can be computed classically by using
linerized Einstein equations and it turns out to equals  the
horizon area. Finally one arrives to the remarkable result\cite{son1}:
\begin{equation}
\frac{\eta}{s}=\frac{1}{4\pi}.
\end{equation}
Curiously enough this result is quite independent  on the
particular form of the metric and it is the same for Dp, M2 and M5 branes 
the reason being the universality of the graviton
absorption cross section.

The above result for $\eta/s $, together with the absence of an empirical counterexample in spite of the many fluids known, was the inspiration for 
Kovtun, Son and Starinets (KSS)  to set
the conjecture that, for
a very wide class of systems, including those that can be
described by a sensible (i.e. ultraviolet complete) quantum field theory,
 the above ratio has the lower bound $1/4\pi$. There are many theoretical arguments which
 favour this bound, both theoretical and experimental, including the plasma produced  
 in heavy ion collisions which seems to be near the saturation of the bound.
 However some controversy has been raised recently about the 
 universal applicability of the bound\cite{status}.

 If correct, one of the most obvious  consequences of the KSS conjecture 
is the absence of perfect fluids in Nature. (In fact this could be welcome since 
perfect fluids are known to give rise to problems such as the d'Alembert paradox). 
More recently Bekenstein and collaborators have pointed out that the accretion of an
ideal fluid onto a black hole could violate the Generalized Second
Law of  Thermodynamics\cite{bekenstein}, suggesting a possible
connection between this law and the KSS bound.

\section {The RHIC case}
During the last years the Relativistic Heavy Ion Collider (RHIC) 
has been producing a large number of Au+Au collisions (A= 197),  at a  center of mass energy
per nucleon of $E=200$ GeV with a luminosity  ${\mathcal{L}}= 2\times
10^{26}$ cm$^{-2}$ s$^{-1}$. From its  four experiments  STAR, PHENIX,
BRAHMS and PHOBOS it has been possible to obtain a lot of experimental information
with important phenomenological consequences. 

First of all thermochemical models
 describe well the different particle yields
fitting  $T=177$ MeV and the baryon chemical potential $\mu_B = 29$ MeV
for  $E_{CM}  = 200$ GeV per nucleon. From the observed transverse and rapidity distributions, the Bjorken
model predicts an energy density at time $t_0 = 1$ fm  of $4$ GeV
fm$^{-3}$,  which strongly suggest that the produced matter may be well above
 the threshold for Quark
Gluon Plasma (QGP) formation. In addition, hydrodynamical models  with very low viscosity
 reproduces the measurements
of radial and elliptic flow up to transverse momenta of 1.5 GeV.
The collective flow is probably generated early in the collision before hadronization. 
The QGP seems to be more strongly interacting than expected on the basis of
perturbative QCD and asymptotic freedom (hence the low viscosity) and some preliminary
 estimations of $\eta/s$ based on elliptic flow\cite{shuryak}$^,$\cite{Teaney} and transverse momentum correlations\cite{gavin} seem to be compatible with  a value close to 0.08 (the
KSS bound). This would make hadronic matter in this regime  the most perfect fluid known.

\section{$\eta/s$ and the phase transition}

Recently Csernai, Kapusta and McLerran\cite{CKM} made the
observation that, in all systems whose $\frac{\eta}{s}(T)$ plot
has been examined, the minimum of $\eta/s$ and
the liquid-gas phase transition seem to happen at the same temperature.\\

Since no general demonstration of this coincidence is known either, we set out to clarify it in a controlled model setup. We chose the large-$N$ Linear Sigma Model (L$\sigma$M)\cite{Usrecent} that is a theory sufficiently ressembling the pion gas that our hadron-physics computer codes to solve Boltzmann's equations can be directly used.  At the same time one knows when and how the phase transition occurs, unlike the case of chiral perturbation theory where one is bound to the low-energy phase.

The model Lagrangian is
\be \label{lagrangian} \mathcal{L} [\Phi, \pa_{\mu} \Phi] = \frac{1}{2} \ \pa_{\mu} \Phi^T \pa^{\mu} \Phi + \ov{\mu}^2 (\Phi^T \Phi)- \lambda (\Phi^T \Phi)^2 \ee
where $\ov{\mu}^2$  is positive (opposite in sign to a scalar field mass term) and $\lambda > 0$. With this choice of parameters the L$\sigma$M presents  spontaneous symmetry breaking from $SO(N+1)$ to $SO(N)$. The field $\Phi$ acquires a vacuum expectation value (VEV) where the field configuration of minimum energy verifies (at tree level)
\be \Phi^T \Phi = \frac{\ov{\mu}^2}{2 \lambda} = f_{\pi}^2 = NF^2.\ee
Denoting $\pi^a \ (a=1,\cdots,N)$ to the $N$ first components of $\Phi$ and $\sigma$ the $N+1$ component we can choose
the VEV in the direction of the latter. Thus we have $\langle \pi^a \rangle=0$ but $\langle \sigma \rangle = f_{\pi}$. The pions are the $N$ massless Nambu-Goldstone bosons; on the other hand, the field $\sigma$ acquires a mass equal to $m_{\sigma}^2=8 \lambda NF^2$.
Taking the limit $m_{\sigma}^2 \rightarrow \infty$ one can express $\sigma$ in terms of the pions as $\sigma=\sqrt{f_{\pi}^2-\pi^a \pi^a}$. This is the non-linear sigma model in which one can eliminate explicitly the $\sigma$ degree of freedom.

 Our first finding is that $\eta/s$ presents a non-analyticity as a function of $T$ at $T=T_c$ where a very different qualitative behaviour of $\eta/s$ begins. This is due to the dependence of $\eta/s$ on the  thermal pion mass $m_\pi(T)$ and the $\sigma$ condensate, both non-analytical at $T_c$. 
Below $T_c$, the thermal pion mass  is identically zero (the Goldstone theorem protects the pion masslessness from radiative corrections). In this phase the symmetry is still broken and the condensate $\langle \sigma \rangle$ is non-zero. Above $T_c$ the symmetry is restored and $\langle \sigma \rangle$ vanishes. The ``classical'' pion mass $m_{\pi}$ is zero as well. However, quantum thermal corrections force a temperature-dependent thermal mass, $m_{\pi} (T)$,
\ba
\langle \sigma(T)\rangle = 
\left\{ \begin{array}{lr}
 f_{\pi} \left( 1- \frac{T^2}{T_c^2} \right)^{1/2} & T < T_c \\
0 & T \ge T_c
\end{array}
\right., 
\\
 m_{\pi}^2(T)= 
\left\{ \begin{array}{lr}
 0 & T < T_c \\
\frac{N}{3} \lambda_R (T^2-T^2_c) & T \ge T_c
\end{array}
\right. \ .
\ea
This mass is continuous at $T_c$ but non-analytic. The behaviour of the condensate $\langle \sigma \rangle$ itself is also continuous  but with a discontinuous derivative at $T_c$. These two quantities influence both $\eta$ and $s$ and their non-analiticity is inherited by the KSS ratio.

By varying $f_\pi$ or $N$, the minimum of $\eta/s$  moves according to the $N$ and $f_{\pi}$ dependences of $T_c$, providing evidence for our claim that $T_c$ and the minimum of $\eta/s$ are related. 

\begin{figure}
\begin{centering}
\includegraphics[height=2.5in]{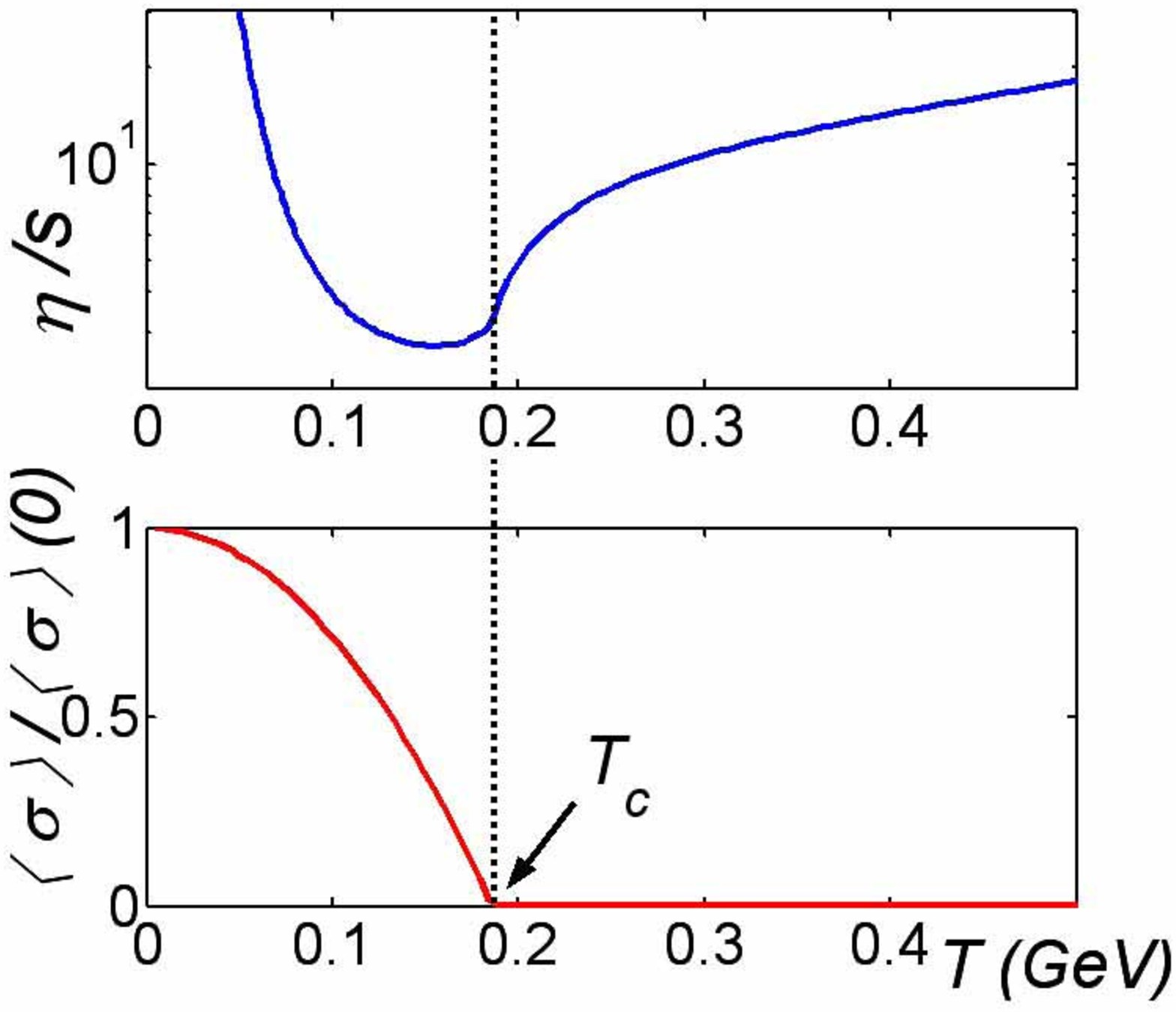}
\caption{The minimum of $\eta/s$ occurs just before the critical temperature for the phase transition in the Linear Sigma Model. This is where the condensate varies rapidly approaching zero. The phase transition is marked by a non--analyticity of the ratio.
\label{fig:minimumTc}}
\end{centering}  
\end{figure}

In Fig.~\ref{fig:minimumTc} we zoom in the $\eta/s$ plot near the critical temperature.
As can be seen, the minimum is not reached at the critical temperature, but right before $T_c$. This shows that the minimum of $\eta/s$ is controlled  by the rapid variation of the order parameter. To understand this result one needs to keep in mind the diffusive nature of the transport in a gas. 
With increasing temperature, the gas particles carry transverse momentum between different parts of the gas more efficiently, and thus increase the shear viscosity. However their interactions hamper transport. As $\langle \sigma\rangle \propto F$ decreases rapidly, the pion elastic cross section increases. Since $\eta\propto 1/\sigma_{\pi\pi}$ in kinetic theory, the viscosity must drop. Eventually growth is regained as the temperature increases.

\section{Outlook}

One could argue that our mean field calculation of the vacuum structure 
could be improved by employing the 2PI formalism to compute the effective action. However we think the result of these, more elaborate computations, would be the same around the physical minimum of the effective potential.
The phase transition is marked by a non-analiticity of $\eta/s$, as for many other thermodynamic quantities, but the minimum does not need to exactly coincide.

A similar result has been obtained for the bulk viscosity~\cite{Danirecent} 
in the $1+1$-dimensional Gross-Neveu model, where the maximum of the 
trace anomaly does not coincide with the maximum of the bulk viscosity, that has just a sharp change of derivative there.

We are looking forward to further computations of transport coefficients in various systems, and to what insight the AdS/CFT conjecture can bring us.

\vskip 1.0cm


\end{document}